\documentclass[conference]{Article}
\IEEEoverridecommandlockouts
\usepackage{cite}
\usepackage{amsmath,amssymb,amsfonts}
\usepackage{algorithmic}
\usepackage{graphicx}
\usepackage{textcomp}
\usepackage{xcolor}
\def\BibTeX{{\rm B\kern-.05em{\sc i\kern-.025em b}\kern-.08em
    T\kern-.1667em\lower.7ex\hbox{E}\kern-.125emX}}
\begin{document}

\title{Artificial Intelligence, VR, AR and Metaverse Technologies for Human Resources Management\\
}

\author{\IEEEauthorblockN{Ömer Aydın}
\IEEEauthorblockA{\textit{Department of Electrical and Electronics Engineering} \\
\textit{Faculty of Engineering and Natural Sciences}\\
\textit{Manisa Celal Bayar University}\\
Manisa, Türkiye \\
\texttt{\{omer.aydin\}@cbu.edu.tr} \\
https://orcid.org/0000-0002-7137-4881
}
\and
\IEEEauthorblockN{Enis Karaarslan}
\IEEEauthorblockA{\textit{Department of Computer Engineering} \\
\textit{Faculty of Engineering}\\
\textit{Mugla Sitki Kocman University}\\
Mugla, Türkiye \\
\texttt{\{enis.karaarslan\}@mu.edu.tr} \\
https://orcid.org/0000-0002-3595-8783
}
\and
\IEEEauthorblockN{Nida Gökçe Narin}
\IEEEauthorblockA{\textit{Department of Statistics} \\
\textit{Faculty of Science}\\
\textit{Mugla Sitki Kocman University}\\
Mugla, Türkiye \\
\texttt{\{gnida\}@mu.edu.tr} \\
https://orcid.org/0000-0002-4840-5408
}

}
\maketitle

\begin{abstract}
Human Resources (HR) technology solutions encompass software and hardware tools designed to automate HR processes, gather, process, and analyze data, utilize it for strategic decision-making, and execute HR professionals' tasks while prioritizing security and privacy considerations. As with numerous other domains, Digital Transformation and emerging technologies have commenced integration into HR processes. These technologies are utilized by HR professionals and various stakeholders involved in HR operations. This study evaluates the utilization of Artificial Intelligence (AI), Virtual Reality (VR), Augmented Reality (VR), and the Metaverse within HR management, focusing on current trends and potential opportunities. A survey was conducted to gauge HR professionals' perceptions and critiques regarding these technologies. Participants were the HR department officers, academicians who specialized in HR and staff who had courses at diverse levels about HR. The acquired results were subjected to comparative analysis within this article.

\end{abstract}


\section{Introduction}
The Metaverse can provide face-to-face interaction that is missing in a dispersed work environment, which is critical to fostering employee connections. Metaverse training can create a memorable, immersive, and captivating learning environment that allows employees to learn and develop at a pace that is conducive to the always-on, hybrid world we live in. The Metaverse can also be used for virtual meetings, recruiting and hiring interviews, onboarding, and learning and development, which can help employees feel more connected to their colleagues and the organization. The benefits of using Metaverse technology in HR are immense. The Metaverse has the potential to improve the organizational climate, provide better accessibility, and make more efficient use of resources. It can be used for virtual meetings, recruiting and hiring interviews, onboarding, employee engagement, and communication regardless of location. The Metaverse can also help HR improve new employee inductions, employment information sharing, online aptitude tests, and AI job capacity evaluation. The Metaverse can be used to enhance employee engagement experiences by providing a space for employees to collaborate, share ideas, and work on projects. It can also be used to train employees and provide them with the opportunity to learn new skills, creating a more engaging and motivating work environment. Other examples of Metaverse technologies for employee engagement include linking company communications with employee engagement, consistent messaging, regular daily/weekly updates, increased executive visibility, and open, two-way communication. Virtual reality training experiences can improve employee performance by 70\%, and employers can train and educate employees by helping them run through scenarios in a virtual environment. However, HR professionals will need to reassess how to educate and train employees in these new technologies and software as they become more prevalent in the industry. Metaverse training can create a memorable, immersive, and captivating learning environment that allows employees to learn and retain information more effectively. The Metaverse can also be used for virtual meetings, onboarding, and learning and development, which can help employees feel more connected to their colleagues and the organization. By leveraging established Metaverse technologies such as avatars, gaming, VR, and hand-tracking controllers, HR leaders can ease employees into the habit of using the Metaverse and make it a more engaging experience. Overall, the Metaverse has the potential to enhance employee engagement by providing a more immersive and interactive work environment.
Our study is conducted to seek answers to some research questions. The research questions of this article are as follows:
\begin{itemize}
\item How are Artificial Intelligence (AI) technologies currently used in Human Resources (HR) management, and what are HR professionals' opinions on the advantages and disadvantages of their implementation?
\item What is the potential use of Virtual Reality (VR) and Augmented Reality (AR) technologies in HR processes, and what are HR professionals' opinions on the advantages and disadvantages of using these technologies in HR processes?
\item What are the emerging trends in the use of the Metaverse in HR management, and what are the possible advantages and disadvantages?
\end{itemize}
The structure of the article is as follows: In the second part, information about HR operations will be shared. In the subsequent section, AI, VR, AR, Metaverse, and HR-related publications in the literature will be examined. Following that, AR, VR, AI, and Metaverse usage in HR operations will be presented in detail under subheadings. This will be followed by a section where information about the research method and the tools used will be shared. Subsequently, the data will be introduced in the subsequent section. The results obtained by processing the data will be shared under the seventh section. In the last section, the results will be discussed, and then the article will be concluded.

\section{HR Operations}
AI applications in HR management include resume scanning, task automation, real-time talent management, reducing employee churn rates, data analysis, recruitment, training, onboarding, performance analysis, and retention. AI can streamline the recruitment process by designing more user-friendly job application forms and identifying the best candidates for a job. AI can also automate low-value HR tasks, such as employee records management and payroll, freeing up HR professionals to focus on more strategic work. AI-powered chatbots can assist with employee inquiries and provide real-time feedback to employees. AI can analyze employee data to provide valuable insights into employee performance, behavior, and engagement, which can help HR teams make better decisions. Overall, AI technologies are transforming HR management by providing valuable insights, automating low-value tasks, and improving the efficiency and effectiveness of HR processes. Some examples of companies using AI in HR workflows include SAP, which uses AI in HR recruitment to streamline the hiring process and automate tasks like resume screening. ServiceNow is another company that uses AI-powered processes to customize background checks, ensure diversity in hiring, train employees, and improve employee engagement. Other companies that are embracing AI in HR management include IBM, Google, and Amazon. These companies use AI to automate low-value HR tasks, such as employee records management and payroll, freeing up HR professionals to focus on more strategic work. They are also using AI to analyze employee data to provide valuable insights into employee performance, behavior, and engagement, which can help HR teams make better decisions.

\section{Literature Review}
Some researchers investigate the impact of information technology on the human resource (HR) function. The use of rapidly evolving information technology has brought significant changes to HR. Haines et al. show that information technology enables HR to be more involved in strategic roles such as being a business partner and a change agent. Additionally, the study suggests that information technology use is associated with increased technical and strategic effectiveness in HR. This indicates the transformative potential of information technology-supported HR management applications. Overall, the study emphasizes the significant influence of information technology on reshaping and enhancing HR function (Haines et al., 2008). Another research discusses the potential impact of workforce management applications on enterprise-wide applications market growth. However, it raises the question of whether firms will genuinely benefit from these applications, considering numerous horror stories about failed technology implementation efforts that erode confidence. To address this issue, the text proposes treating the installation of HR technology as a form of innovation. It introduces a model that describes the technology implementation process and highlights various issues that academics and practitioners should consider (Shrivastava \& Shaw, 2003). It is one of the popular concepts of today's digital transformation. Some studies aim to examine the evolving role of human resource management in the context of digital transformation. Digital transformation involves leveraging new technology to transform business processes, operations and structures for competitive advantage. Some researchers consider human capital, intellectual capital and knowledge as critical components in achieving this advantage. It shows that digitalization affects daily HR practices, especially with the use of human resources information systems (Fenech et al., 2019).
Some articles discuss the impact of AI and robotics on various industries. Gulliford and Dixon focused on organizational science and workforce data analysis conducted by Qlearsite to discuss the impacts of these technologies. The paper tracks the evolution of AI implementation, detailing how businesses can leverage AI to understand their workforce better, identify performance barriers, and enhance productivity. Despite initial challenges faced by senior HR members, the article highlights the tangible benefits of AI implementation in improving corporate success and productivity. It emphasizes AI's ability to solve complex issues, such as absenteeism, through data analysis, which proves to be invaluable for HR leaders and senior decision-makers (Gulliford \& Dixon, 2019). Michailidis discusses the impact of technological innovations, such as blockchain and artificial intelligence (AI), on human resource (HR) practices within business and non-profit organizations. It highlights the influence of these innovations on employment patterns, hiring practices, and employee management. The paper is structured into three main sections: the first examines the effects of blockchain and AI on HR practices, the second discusses hiring practices at firms, and the third analyzes employment patterns in the age of high-tech super-automation. Additionally, there is a concluding section that explores the implications of AI on employment and income inequality in society (Michailidis, 2018). Pandey and Khaskel’s study explores the challenges AI presents to HR functions and the technological solutions adopted by companies like IBM and the UnitedHealth Group. With over 50\% of the working population being Gen Y (under 25) and around 65\% under 35, their reaction to HR transformation is crucial. A qualitative study involving interviews with Gen Y professionals identified HR themes suitable for AI applications. Results show a clear majority supporting AI implementation in HR systems, indicating readiness among the working population(Pandey \& Khaskel, 2019). The paper, authored by Nyathani in 2023, establishes the foundational principles underpinning effective HR analytics, accentuating the paramount importance of meticulous employee data management encompassing collection, storage, quality assurance, and governance across diverse employee data types. Nyathani's work highlights the central role of AI in transforming HR analytics, demonstrating its capacity to augment the processing and analysis of employee data. Ethical considerations, including bias mitigation, data privacy, and responsible AI governance, are rigorously addressed by Nyathani, alongside a forward-looking discussion on emerging trends and the evolving role of HR professionals within an AI-driven HR landscape (Nyathani, 2023).
AR and VR technologies are poised to improve the onboarding process by providing virtual office tours, eliminating the need for new hires to be physically present. Likewise, the utilization of virtual and digital learning platforms for reskilling and upskilling employees is set to become prevalent, negating the necessity for organizations to invest in the infrastructure and logistics needed for traditional hands-on training (Am et al., 2020). Yawson expressed the notion that emerging technologies such as Artificial Intelligence (AI), Virtual Reality (VR), Augmented Reality (AR), and Blockchain, along with data science, are driving the automation and digitization of HR functions. These advancements have the potential to facilitate decision-making processes with reduced implicit bias, thereby establishing a foundation for a more equitable decision-making process throughout the organization (Yawson, 2020).
Vochozka et al. (2022) emphasize the scarcity of research on immersive workspaces within the fully connected Metaverse. Their article compiles previous research indicating that augmented reality technologies, simulation modeling software tools, and workplace tracking systems facilitate immersive work environments. In their quantitative literature review conducted throughout May 2022, they identified 157 articles meeting eligibility criteria from databases such as Web of Science, Scopus, and ProQuest. Through meticulous scrutiny, they refined their dataset to 31 generally empirical sources, ensuring the reliability and relevance of their findings. Lim et al. (2023) conducted a study examining the potential applications of the Metaverse in various aspects of human resource development (HRD) and assessed its significance for both employee and organizational growth. Utilizing a narrative review approach, the authors meticulously examined 34 cases, uncovering several noteworthy findings that contribute significantly to the HRD domain. Firstly, the study sheds light on a critical emerging trend overlooked in the HRD field, urging researchers and practitioners to pay closer attention. Secondly, it offers an analytical perspective on the Metaverse in comparison to other training technologies, evaluating its efficacy in fostering employee development. Thirdly, Lim et al.'s research enriches the understanding of HRD professionals regarding Metaverse-based interventions, providing practical and theoretical insights into how such interventions can reshape organizational culture and enhance employee performance. Finally, the study addresses pertinent theoretical and practical inquiries concerning HRD challenges stemming from the Metaverse.

\section{AR, VR, AI and Metaverse Usage in HR Operations}
VR and AR are two emerging technologies that have the potential to revolutionize the way HR is conducted. VR creates a fully immersive experience that allows users to interact with a virtual world, while AR overlays digital information onto the real world. Both technologies hold promise for various HR applications, including recruiting and onboarding, training and development, employee engagement, performance management, wellness, and safety. Additionally, AI enables talent acquisition through automated screening processes that efficiently filter through large volumes of resumes to identify top candidates. AI-powered chatbots further streamline employee queries and support, providing instant assistance and enhancing the overall employee experience. Moreover, predictive analytics driven by AI algorithms assist HR professionals in forecasting workforce trends, facilitating proactive strategies for talent retention and succession planning. Beyond VR, AR, and AI, the integration of the Metaverse into Human Resources (HR) signifies a profound shift in organizational workforce management. The Metaverse offers immersive virtual environments where HR professionals can engage with employees and candidates in novel ways. Now, let's briefly explore the relationship between these technologies and HR under the following subheadings.

\subsection{AI in HR}
AI is being used in HR to automate tasks, improve decision-making, and provide insights. This is freeing up HR professionals to focus on more strategic tasks, such as talent development and employee engagement.
AI is being used in HR in recruiting and hiring, talent management, employee development and employee engagement. AI can be used to screen resumes, conduct interviews, and even make hiring decisions. AI can be used to analyze employee data to identify trends and patterns that can help HR professionals make better decisions about things like compensation, benefits, and training. AI can be used to create personalized learning plans for employees, track their progress, and provide feedback. AI can be used to track employee sentiment, identify areas of improvement, and provide personalized feedback. The details will be given in the following subsections.

\subsubsection{Recruiting and hiring}
AI can be used to screen resumes by identifying keywords and phrases that are relevant to the job opening. This can help HR professionals quickly and easily identify qualified candidates. AI can also conduct interviews by asking questions and evaluating the candidate's responses. This can help HR professionals to get a better sense of the candidate's skills and abilities. Finally, AI can even be used to make hiring decisions by scoring candidates and recommending the top candidates for the job (Li et al., 2021).

\subsubsection{Talent management}
AI can be used to analyze employee data to identify trends and patterns that can help HR professionals make better decisions about things like compensation, benefits, and training(Charlwood, 2021). For example, AI can be used to analyze employee performance data to identify areas where employees need improvement. This information can then be used to develop training programs that can help employees improve their performance. AI can also be used to analyze employee satisfaction data to identify areas where employees are unhappy. This information can then be used to make changes to the workplace that can improve employee satisfaction(Rath et al., 2023).

\subsubsection{Employee development} 
AI can be used to create personalized learning plans for employees, track their progress, and provide feedback. This can help employees develop the skills they need to succeed in their careers (Eubanks, 2022). For example, AI can be used to create learning plans that are tailored to each employee's individual needs and interests. AI can also be used to track employee progress and provide feedback on their learning. This can help employees stay motivated and on track to achieve their learning goals.

\subsubsection{Employee engagement} 
AI can be used to track employee sentiment, identify areas of improvement, and provide personalized feedback(Mer and Srivastava, 2023). This can help HR professionals create a more positive and productive work environment for their employees. For example, AI can track employee sentiment through surveys and polls. This information can then be used to identify areas where employees are unhappy. HR professionals can then work to address these issues and create a more positive work environment for their employees.

\subsection{VR/AR in HR}
Potential benefits of using VR and AR in HR are reduced costs, improved productivity and increased employee satisfaction.
VR and AR can help to reduce costs in several ways. For example, VR can be used to create virtual job simulations that eliminate the need for travel and accommodation for candidates. AR can be used to provide employees with training on new products or services without the need for costly classroom training.
VR and AR can help to improve productivity in many ways. For example, VR can be used to create immersive training simulations that allow employees to learn new skills more quickly and effectively. AR can be used to provide employees with real-time information and instructions that can help them to perform their jobs more efficiently.
VR and AR can help to increase employee satisfaction in several ways. For example, VR can be used to create interactive and engaging experiences that can help to improve employee morale and motivation. AR can be used to provide employees with a more personalized and supportive work environment.
 
\subsection{Metaverse and HR}
Beyond affecting a limited area (Bale et al., 2022; Henz, 2022), Metaverse is preparing to significantly impact many professional business areas, such as the human resources (HR) field. In the Metaverse environment, individuals can interact with digital representations of themselves and others. This way, one can participate in immersive experiences and create or consume content in virtual environments.
As organizations explore the potential of metadata, HR professionals face new challenges and opportunities in talent acquisition, training and employee engagement. One of the primary impacts of Metaverse for HR is its potential to redefine traditional concepts of workplace and work culture (Zvarikova et al., 2022). As virtual offices and remote working become increasingly common, HR departments need to adapt their strategies to foster a sense of belonging and community among geographically dispersed but digitally connected employees (Bennett, 2022). Additionally, HR policies and practices will need to address issues such as digital identity, privacy, and security in the Metaverse environment and ensure that employees feel comfortable and safe in their virtual interactions(Di Pietro \& Cresci, 2021; Wang et al., 2022; Far \& Rad, 2022; Zhao et al, 2023). Moreover, the Metaverse also offers new opportunities for talent acquisition and hiring. Companies can use virtual job fairs, immersive experiences, and gamified assessments to attract and evaluate candidates from diverse backgrounds and locations. Virtual reality (VR) simulations can also be used for realistic job previews and skills assessments, and have the potential to provide a more accurate representation of a candidate's abilities and suitability for the role.
In terms of employee training and development, Metaverse offers a rich environment for immersive learning experiences. VR-based training programs simulate real-world scenarios, allowing employees to practice skills and decision-making in a risk-free environment(Hawkins, 2022). Additionally, virtual collaboration tools and platforms enable employees to collaborate, learn from each other, and share knowledge across geographic boundaries, fostering a culture of continuous learning and innovation.
However, as organizations embrace the Metaverse, it is important to be mindful of potential challenges and ethical considerations(Rozak et al, 2023). Issues such as digital equity, accessibility and inclusion must be addressed to ensure all employees have equal opportunities to participate and thrive in virtual environments. HR professionals will play a pivotal role in promoting diversity, equity and inclusion in the Metaverse, and they must take part in creating welcoming and accessible virtual spaces for everyone. To summarize, Metaverse offers a paradigm shift in the way HR works, opening up new possibilities for talent management, employee development and organizational culture. By embracing emerging technologies and reimagining traditional practices, HR departments can leverage the full potential of the Metaverse to create engaging, inclusive, and future-ready workplaces.

\section{Method}
A survey was conducted involving 158 participants comprising academicians specializing in HR-related topics, HR professionals, HR officers, and individuals who have undergone HR-related courses. The survey aimed to gather insights into the utilization of emerging technologies, namely Virtual Reality (VR), Augmented Reality (AR), Artificial Intelligence (AI), and the integration of the Metaverse, within the field of HR.
To execute the survey, Google Forms was utilized to create the questionnaire, and invitations were distributed via email to the specified participant groups. Participants were asked to respond to various questions regarding their perceptions and critiques of these technologies in HR management.
Upon completion of the survey, the responses were collected and compiled into an Excel file for data management. Preprocessing of the data was conducted, which involved grouping certain responses and analyzing and processing responses provided under the "Other" option.
Subsequently, the processed data were subjected to statistical analysis using the Jamovi software, a powerful R-based tool for statistical analyses. Jamovi was selected for its user-friendly interface and robust analytical capabilities. Various statistical tests and analyses were performed using Jamovi (Edelsbrunner, 2017; Jamovi, 2024) to derive insights and draw conclusions from the survey data.
The utilization of Jamovi facilitated the examination of relationships, trends, and patterns within the dataset, allowing for a comprehensive evaluation of the survey results (Şahin \& Aybek 2019; Bartlett \& Charles, 2022). The statistical findings obtained through Jamovi were instrumental in informing the discussion and conclusions of this study regarding the application and implications of emerging technologies in HR management.

\section{Data}
The data file includes participants' demographic characteristics (age, gender, educational level), professional experience (time working in the field of human resources), and their opinions on the use of these new technologies. Participants' responses reflect their perceptions and expectations about the advantages and disadvantages of artificial intelligence, virtual reality, augmented reality and Metaverse technologies in human resources processes.
The questions in the survey and the answer options that can be given to these questions are as follows:

\textit{Demographic Information: }
\begin{itemize}
    \item Country
    \item Age
    \item less than 20, 20-30, 31-40, 41-50, 51-60, over 60
    \item Gender
    \item (Male, Female)
    \item Educational Status
    \item Secondary School or lower education level, High School, University (Associate Degree), University (Bachelor's Degree), University (Master's Degree), University (PhD)
\end{itemize}

\textit{Experience in Human Resources:}
\begin{itemize}
    \item What is the basis of human resources expertise?
    
    (Working in the HR unit, undergraduate education related to HR, received training/certificate related to HR, other)
    \item Professional Experience 
    
    (0-1 Years, 1-3 Years, 3-5 Years, 5-10 Years, 10-20 Years, More than 20 Years)
\end{itemize}

\textit{Evaluating the Use of Artificial Intelligence (AI):} 
\begin{itemize}
    \item Advantages of using artificial intelligence (AI) technology in Human Resources processes? 
    
    (Increasing productivity, improving the recruitment process, supporting training and development, facilitating performance management, no advantage, Other)
    \item Disadvantages of using artificial intelligence (AI) technology in Human Resources processes? 
    
    (Reduced human interaction, risk of producing erroneous results, ethical issues, data security risks, will cause employees to lose their jobs, makes employees accustomed to laziness, no disadvantage, other)
\end{itemize}

\textit{Evaluation of Virtual Reality (VR) Usage:}

\begin{itemize}
    \item Benefits of using virtual reality (VR) technology in Human Resources management?
    
    (Improving training and simulation experiences, facilitating remote collaboration and communication, improving recruitment and interview processes, increasing employee loyalty, no benefit, other)
    \item Benefits of using virtual reality (VR) technology in Human Resources management?
    
    (High costs, risk of physical discomfort, difficulty in the acceptance process, data security risks, no disadvantages, other)
\end{itemize}

\textit{Evaluation of Augmented Reality (AR) Usage: }

\begin{itemize}
    \item Benefits of using augmented reality (AR) technology in Human Resources management?
    
    (Better candidate evaluation during the recruitment process, enriching training and development programs, increasing the work efficiency of employees, improving workplace safety and risk management, no advantage, other)
    \item Disadvantages of using Augmented Reality (AR) technology in Human Resources processes?
    
    (Ensuring adequate infrastructure and technical requirements, high operating costs, data security risks, employee resistance or difficulty in the acceptance process, no disadvantages, other)
\end{itemize}

\textit{Evaluation of Metaverse Usage:}

\begin{itemize}
    \item Benefits of using Metaverse technology in Human Resources Management?
    
    (Global collaboration and remote working opportunities, creative and interactive training experiences, increasing employee participation and motivation, improving recruitment and interview processes, none, other)
    \item Disadvantages of using Metaverse technology in Human Resources processes?
    
    (Complexity of technological infrastructure, decrease in human interaction, data privacy and security concerns, risk of people disconnecting from reality, possibility of people experiencing psychological problems, no weaknesses, other)
\end{itemize}

A total of 158 participants participated in the survey. When we examine the data obtained as a result of the survey, the frequencies are as follows. 
\begin{table}[]
\caption{Frequency of Survey Language}
\resizebox{\columnwidth}{!}{%
\begin{tabular}{|c|c|c|c|}
\hline
\multicolumn{1}{|l|}{\textbf{Survey Language}} & \multicolumn{1}{l|}{\textbf{Counts}} & \multicolumn{1}{l|}{\textbf{\% of Total}} & \multicolumn{1}{l|}{\textbf{Cumulative \%}} \\ \hline
English& 3& 1.9\%& 1.9\% \\ \hline
Turkish& 155 & 98.1\%  & 100.0\% \\ \hline
\end{tabular}
}
\end{table}
The survey was applied with forms created in two different languages. In the survey form prepared with Google Forms, the participant was first asked about his/her language preference and then the questions in the relevant language were shown. The questions asked and the answer options offered are the same in both languages. In Table I, we can see in which language the survey participants answered the survey questions. Only 3 of the participants answered the survey by choosing the English language. 155 participants preferred to answer the questions in Turkish.

\begin{table}[]
\caption{Frequency of Country}
\resizebox{\columnwidth}{!}{%
\begin{tabular}{|c|c|c|c|}
\hline
\multicolumn{1}{|l|}{\textbf{Country}} & \multicolumn{1}{l|}{\textbf{Counts}} & \multicolumn{1}{l|}{\textbf{\% of Total}} & \multicolumn{1}{l|}{\textbf{Cumulative \%}} \\ \hline
India& 3& 1.9\%& 1.9\% \\ \hline
Kyrgyzstan& 1 & 0.6\%  & 2.5\% \\ \hline
Türkiye& 154 & 97.5\%  & 100.0\% \\ \hline
\end{tabular}
}
\end{table}
The survey was shared in different environments so that the participants were from different geographies and countries. In addition to inviting participants via email, posts were also shared on Linkedin and other social media platforms, and invitations were sent to people. Table II shows which countries participated in the survey. 1 of the participants is from Kyrgyzstan, 3 are from India and 154 are from Türkiye.

\begin{table}[]
\caption{Frequency of Age}
\resizebox{\columnwidth}{!}{%
\begin{tabular}{|c|c|c|c|}
\hline
\multicolumn{1}{|l|}{\textbf{Age}} & \multicolumn{1}{l|}{\textbf{Counts}} & \multicolumn{1}{l|}{\textbf{\% of Total}} & \multicolumn{1}{l|}{\textbf{Cumulative \%}} \\ \hline
less than 20&4 & 2.5\%  & 2.5\% \\ \hline
20-30       &34& 21.5\% & 24.1\% \\ \hline
31-40       &44& 27.8\% & 51.9\% \\ \hline
41-50       &54& 34.2\% & 86.1\% \\ \hline
51-60       &17& 10.8\% & 96.8\% \\ \hline
over 60     &5 & 3.2\%  & 100.0\% \\ \hline
\end{tabular}
}
\end{table}
The ages of the participants in our survey is an issue that we care about in our evaluations. In particular, determining how participants in different age groups will answer the relevant survey questions will enable us to make important inferences. Table III shows the age groups of the participants.

\begin{table}[]
\caption{Frequency of Gender}
\resizebox{\columnwidth}{!}{%
\begin{tabular}{|c|c|c|c|}
\hline
\multicolumn{1}{|l|}{\textbf{Gender}} & \multicolumn{1}{l|}{\textbf{Counts}} & \multicolumn{1}{l|}{\textbf{\% of Total}} & \multicolumn{1}{l|}{\textbf{Cumulative \%}} \\ \hline
Female& 81& 51.3\%& 51.3\% \\ \hline
Male& 77 & 48.7\%  & 100.0\% \\ \hline
\end{tabular}
}
\end{table}
Table IV shows the distribution of participants by gender. It can be said that the distribution is balanced. 51.3\% of the participants are women and 48.7\% are men.
\begin{table}[]
\caption{Frequency of Education Level}
\resizebox{\columnwidth}{!}{%
\begin{tabular}{|c|c|c|c|}
\hline
\multicolumn{1}{|l|}{\textbf{Education Level}} & \multicolumn{1}{l|}{\textbf{Counts}} & \multicolumn{1}{l|}{\textbf{\% of Total}} & \multicolumn{1}{l|}{\textbf{Cumulative \%}} \\ \hline
Secondary School or lower education level& 0& 0.0\%& 0.0\% \\ \hline
High School& 3 & 1.9\%  & 1.9\% \\ \hline
University (PhD)& 91 & 57.6\%  & 59.5\% \\ \hline
University (Master's Degree)& 28 & 17.7\%  & 77.2\% \\ \hline
University (Bachelor's Degree)& 30 & 19.0\%  & 96.2\% \\ \hline
University (Associate Degree)& 6 & 3.8\%  & 100.0\% \\ \hline

\end{tabular}
}
\end{table}

Another question asked to the participants in the survey is the education level of the participants. Table V shows the distribution of participants according to their education levels. The highest option was "\textit{University (PhD)}", chosen by 97 participants. This is followed by "\textit{University (Bachelor's Degree)}" with 30 participants and "\textit{University (Master's Degree)}" with 28 participants.

\begin{table}[]
\caption{Frequencies of \textit{“I work/have worked in a unit related to human resources”}}
\resizebox{\columnwidth}{!}{%
\begin{tabular}{|c|c|c|c|}
\hline
\multicolumn{1}{|l|}{\textbf{I work/have worked in a unit related to human resources}} & \multicolumn{1}{l|}{\textbf{Counts}} & \multicolumn{1}{l|}{\textbf{\% of Total}} & \multicolumn{1}{l|}{\textbf{Cumulative \%}} \\ \hline
Yes& 53& 33.5\%& 33.5\% \\ \hline
No& 105 & 66.5\%  & 100.0\% \\ \hline
\end{tabular}
}
\end{table}
The working status of the participants in HR-related units is given in Table VI. When the table is examined, it is seen that 33.5\% of the participants work in HR-related units, while 66.5\% do not work in relevant units.
The results regarding the participants' graduation from an undergraduate degree related to HR are shared in Table VII. While 21.5\% of the participants declared that they graduated from HR-related undergraduate departments, 78.5\% did not graduate from HR-related departments.
\begin{table}[]
\caption{Frequencies of “\textit{I received my undergraduate education in a department related to HR}”}
\resizebox{\columnwidth}{!}{%
\begin{tabular}{|c|c|c|c|}
\hline
\multicolumn{1}{|l|}{\textbf{I received my undergraduate education in a department related to HR}} & \multicolumn{1}{l|}{\textbf{Counts}} & \multicolumn{1}{l|}{\textbf{\% of Total}} & \multicolumn{1}{l|}{\textbf{Cumulative \%}} \\ \hline
Yes& 34& 21.5\%& 21.5\% \\ \hline
No& 124 & 78.5\%  & 100.0\% \\ \hline
\end{tabular}
}
\end{table}
In addition, the survey also asked whether the participants had received a certificate or training related to HR. Table VIII contains information regarding this. While approximately 36\% of the participants declared that they had received HR-related training/certificate, approximately 64\% declared that they did not have HR-related training or certification.
\begin{table}[]
\caption{Frequencies of "\textit{I received training/certificate on this subject}"}
\resizebox{\columnwidth}{!}{%
\begin{tabular}{|c|c|c|c|}
\hline
\multicolumn{1}{|l|}{\textbf{I received training/certificate on this subject}} & \multicolumn{1}{l|}{\textbf{Counts}} & \multicolumn{1}{l|}{\textbf{\% of Total}} & \multicolumn{1}{l|}{\textbf{Cumulative \%}} \\ \hline
Yes& 57& 36.1\%& 36.1\% \\ \hline
No& 101 & 63.9\%  & 100.0\% \\ \hline
\end{tabular}
}
\end{table}

Tables VI, VII and VIII attempt to understand how the participants obtained their information about HR. Apart from the options in these tables, some participants declared where they got their information about HR by choosing the "\textit{Other}" option. This information is shared in Table IX. Among those who chose the "\textit{Other}" option, the options "\textit{I teach or am an academic in this field}" and "\textit{I have no expertise}" stand out with the highest rates.
\begin{table}[]
\caption{Frequencies of "\textit{Other}" option}
\resizebox{\columnwidth}{!}{%
\begin{tabular}{|c|c|c|c|}
\hline
\multicolumn{1}{|l|}{\textbf{Other}} & \multicolumn{1}{l|}{\textbf{Counts}} & \multicolumn{1}{l|}{\textbf{\% of Total}} & \multicolumn{1}{l|}{\textbf{Cumulative \%}} \\ \hline
I teach or am an academic in this field& 16& 10.1\%& 10.1\% \\ \hline
I Have General Knowledge & 1 & 0.6\%  & 10.8\% \\ \hline
Not selected this option & 127 & 80.4\%  & 91.1\% \\ \hline
I am working in the field of HR.  & 1 & 0.6\%  & 91.8\% \\ \hline
Insurance and Actuarial & 1 & 0.6\%  & 92.4\% \\ \hline
I have no expertise & 11 & 7.0\%  & 99.4\% \\ \hline
I will work & 1 & 0.6\%  & 100.0\% \\ \hline
\end{tabular}
}
\end{table}

The participants' HR-related experiences in terms of time were also questioned in the survey. In this regard, the participants were asked about the duration of their HR experience. The results are shared in Table X. While 30.4\% of the participants had 10-20 years of experience, the rate of participants with more than 20 years of experience was 24.1\%, and the rate of participants with 5-10 years of experience was 19.6\%. 13.9\% of the participants declared that they had 0-1 year of experience.

\begin{table}[]
\caption{Frequencies of Professional Experience?}
\resizebox{\columnwidth}{!}{%
\begin{tabular}{|c|c|c|c|}
\hline
\multicolumn{1}{|l|}{Professional Experience (Time)} & \multicolumn{1}{l|}{\textbf{Counts}} & \multicolumn{1}{l|}{\textbf{\% of Total}} & \multicolumn{1}{l|}{\textbf{Cumulative \%}} \\ \hline
0-1 Year & 22& 13.9\%& 13.9\% \\ \hline
1-3 Years & 11 & 7.0\%  & 20.9\% \\ \hline
3-5 Years & 8 & 5.1\%  & 26.0\% \\ \hline
5-10 Years  & 31 & 19.6\%  & 45.6\% \\ \hline
10-20 Years & 48 & 30.4\%  & 76.0\% \\ \hline
More than 20 years & 38 & 24.1\%  & 100.0\% \\ \hline
\end{tabular}
}
\end{table}

\section{Results}
The survey results on the utilization of AI, VR, AR, and Metaverse technologies in HR methods are outlined in this section.
As a result of the survey, the rate of those who think that using AI technologies in HR processes will be beneficial was 84.8\% (A\footnote{A: I agree}: 50.6\%; TA\footnote{TA: I totally agree}: 34.2\%). The rate of people who think that there will be no advantage is 11.4\% (D\footnote{D: I disagree}:5.7\%; SD\footnote{SD: Strongly disagree}:5.7\%). Similarly, when the respondents were asked about their opinions on the disadvantages of using Artificial Intelligence technologies in HR processes, 14.6\% stated that they had no opinion, while 51.9\% stated that they had a disadvantage (A:44.3\%; TA: 7.6\%). The rate of those who stated that there was no disadvantage was 33.5\% (D: 25.3\%; SD: 8.2\%). 
These results show that the participants were able to consider together the advantages and disadvantages of the use of AI technologies in HR management. The rate of those who think that using AI technologies in HR management will increase efficiency, improve the recruitment process, support training and development, and facilitate performance management is over 70\%. Ethical problems were determined as the most crucial disadvantage, with 58.9\%. It was observed that the risk of data security came next at 55.1\%, and the risk of producing erroneous results came at 49.4\%.
65.8\% of the participants believe that using VR technology in HR processes would be advantageous (A: 39.2\%; TA: 26.6\%), while 13.3\% (D:7.6\%: SD:5.7\%) thought that it would not have any advantage. Similarly, the rate of those who participated in the survey stating that there would be disadvantages in using virtual reality technology in HR processes was 39.2\% (A: 33.5\%; TA: 5.7\%). 27.8\% of the participants did not express a clear opinion or were undecided. It is seen that the most important benefit of using VR technology in HR management is to improve training and simulation experiences, with 77.8\%. While the rate of those who think it will facilitate remote collaboration and communication is 69.6\%, the rate of those who believe it will improve the recruitment and interview processes is 57.6\%. 84.8\% of the participants think that the use of virtual reality will not increase employee addiction. As with AI technologies, the most significant disadvantage of using VR was the data security risk (50.6\%). Other concerns include high costs (48.7\%), difficulty in the acceptance process (32.9\%), and risk of physical discomfort (24.7\%).
 While the rate of those who think that using AR technology in HR processes would be advantageous is 65.8\% (A: 38.6\%; TA:27.2\%), the rate of those who think that the use of AR will not be beneficial is 14,6\% (D:8.9\%; SD:5.7\%). Similarly, when the respondents were asked whether there would be disadvantages of using Augmented Reality technology in HR processes, 30.4\% remained undecided, while the rate of those who stated that it would have disadvantages was 38.9\% (A: 32.3\%; TA: 6.3\%). Enriching training and development programs with AR was determined as the most important benefit, with 70.9\%. While the rate of those who think it will improve candidate evaluation during the recruitment process is 57.0\%, the rate of those who believe it will improve workplace safety and risk management is 42.4\%. However, 58.9\% of the respondents stated that AR would not increase employees' internal efficiency. The most significant disadvantage of using AR technologies in HR management is high operating costs (53.2\%). This is followed by concerns regarding data security risk (49.4\%), provision of adequate infrastructure and technical requirements (44.9\%), employee resistance, and difficulty of the acceptance process (43.7\%).
While the rate of those who stated that it would be beneficial to use Metaverse technology in HR processes was 53.8\% (A: 32.9\%; TA: 20.9\%), the rate of those who said that it would have disadvantages was 36.5\% (A: 30.4\%; TA:5.1\%).
The most significant benefits of using Metaverse technologies in HR management include global collaboration and remote working opportunities (67.7\%) and creative and interactive educational experiences (61.4\%). The rate of those who think it will increase employee participation and motivation is 43.0\%. 60.8\% of the participants believe it will not improve the recruitment and interview processes. The most significant disadvantage of using Metaverse technology in HR management was determined as the risk of disconnecting people from reality with 54.4\%. While half of the participants saw Metaverse technology's reduction in human interaction and data privacy and security concerns as a disadvantage, the other half did not consider them as such. The rate of those who indicated that it could cause people to experience psychological problems was 41.8\%.
Chi-square analysis (Howell, 2011), which is frequently used in the analysis of categorical data, investigated whether there was a significant difference between the participants' characteristics such as age, gender, professional experience, and education level and their opinions about the use of AI, AR, VR and Metaverse technologies in HR processes. The following results were obtained, with a p-value < 0.05 indicating a statistically significant difference;
\begin{itemize}
\item There is a significant difference of opinion according to gender that AI technology will contribute to improving training and simulation experiences in HR processes (p = 0.023).
\item There is a significant difference of opinion according to gender regarding whether Metaverse technology is a strength in its use in HR management (p=0.004). While 94\% of women think they have strengths, 78\% of men say they have strengths.
\item There is a significant difference in opinion regarding the possibility of producing erroneous results due to the use of AI technologies in HR processes, depending on the level of education (p = 0.0024). 41\% of PhD graduates, 53\% of undergraduate graduates, and 75\% of master's graduates think it can produce erroneous results.
\item It was observed that there was a significant difference in the advantages of using AI technologies in HR processes according to the educational level of the participants (p = 0.02). 91\% of postgraduate graduates stated that they thought it was advantageous.
\item It was observed that there was a significant difference in the disadvantages of using artificial intelligence technologies in HR processes according to educational level (p = 0.024). While 55\% of postgraduate graduates stated that using artificial intelligence technologies in HR processes has disadvantages, 33\% disagreed with this opinion.
\item While 77\% of graduate graduates think that the use of AI technologies in HR processes will support training and development, 77\% of high school and associate degree graduates expressed an utterly negative opinion, which was found to be statistically significant (p = 0.003).
\item There is a statistically significant difference of opinion (0.026) that using Metaverse technologies in HR processes will reduce human interaction. While 55\% of doctoral graduates and all high school graduates expressed the opinion that it would not reduce human interaction, 60\% of undergraduate graduates, 50\% of master's graduates, and all associate degree graduates stated that they thought it would.
\item Opinions that the use of VR technologies in HR processes will improve the training and simulation experience show a significant difference according to gender (p = 0.023). Seventy percent of men and 85 percent of women stated that they would make a positive contribution. In addition, all of the women participating in the survey think that the use of virtual reality technologies will benefit them during their HR hours.
\item 40\% of men and 60\% of women think that Virtual reality technologies pose a data security risk, and the differences in opinion according to gender are statistically significant (p = 0.011).
\item 22\% of men and only 6\% of women stated that using Metaverse technologies in HR processes is not a strong aspect (p=0.004).
\item There is a significant difference between opinions on whether using AR technologies in HR processes will cause physical discomfort depending on whether the participant works in an HR-related unit (p = 0.017). While 87\% of those who work in an HR-related unit state that there is no risk of physical discomfort, 70\% of those who do not work in an HR-related unit think there is no risk.
\item 76\% of the participants who received their undergraduate education from a department related to HR think that the use of AI technologies in the HR process will cause ethical problems (p = 0.019).
\item 50\% of the participants who received their undergraduate education from a department related to HR think these technologies may cause employees to lose their jobs. However, 73\% of the other participants stated that there was no such risk (p = 0.013).
\item While 41\% of the participants who received their undergraduate education from a department related to HR thought that using VR technologies in HR processes involved the risk of physical discomfort, 80\% of the remaining participants stated that there was no such risk (p = 0.012).
\item While 53\% of the participants who received their undergraduate education from a department related to HR thought that the use of AR technologies in HR processes would contribute to enriching training and development programs, 24\% of the remaining participants stated that it did not have such a contribution (p = 0.009).
\end{itemize}

\section{Discussion and Conclusion}
When the participants in the study are evaluated, it is seen that the majority of them have more than 5 years of professional experience. This indicates that the participants consist of experienced individuals in the field. When the education levels of the participants are examined, it is seen that the majority of them hold undergraduate, graduate, and doctoral degrees. There is an almost equal distribution in terms of gender. Regarding age groups, the majority of the participants fall between the ages of 20 and 50. When the study results were analyzed, it was observed that the rate of those who believed that the use of artificial intelligence technologies in HR processes would be beneficial was very high. Similarly, when the participants were asked about their opinions on the disadvantages of using Artificial Intelligence technologies in HR processes, it was observed that half of them thought there would be disadvantages. Approximately one-third of the total participants stated that artificial intelligence has no disadvantages. These results indicate that participants were able to evaluate the advantages and disadvantages of using artificial intelligence technologies in HR management collectively. Those who believe that the use of artificial intelligence technologies in HR management will increase efficiency, improve the recruitment process, support training and development, and facilitate performance management are predominant. Ethical problems were seen as the most significant disadvantage, followed by data security risks and the risk of producing incorrect results.
Concerning virtual reality, it was observed that more than half of the participants expressed the opinion that its use would be advantageous. Approximately one in ten participants thought it would not provide any advantage. Around 40\% of the participants believed that the use of VR in human resources processes would have disadvantages, while one-third of the participants were undecided on this question. It was stated that the most important benefit of this technology is to improve the training and simulation experience. Additionally, many people believe that remote collaboration and communication opportunities will become easier. More than half of the participants stated that the use of VR would benefit the recruitment and interview processes. Security risks, high costs, difficulty in the acceptance process, and risks of physical discomfort were cited as the disadvantages of VR technology.
The number of people who think that the use of augmented reality technology in HR processes will be advantageous is approximately 65\%. The rate of those who think it will not be beneficial remains low. It has been determined that the most important benefit of AR is the enrichment of training and development programs with AR. More than half of the participants think that it will improve candidate evaluation during the recruitment process. Those who believe that it will improve occupational safety and risk management comprise nearly half of the participants. More than half of the participants stated that AR would not increase employees' internal productivity. The most important disadvantage of using AR technologies in HR management was chosen as high operating costs, followed by data security risks, provision of adequate infrastructure and technical requirements, employee resistance, and difficulty in the acceptance process.
More than half of the participants stated that using Metaverse technology in HR processes would be beneficial. One in every three participants stated that using Metaverse in HR processes would have disadvantages. Global collaboration and remote work opportunities, as well as creative and interactive training experiences, were singled out as the most important benefits of using Metaverse technologies in HR management. The rate of those who think that it will increase employee participation and motivation is around 40\%. 60.8\% of respondents think this will not improve recruitment and interview processes. The most important disadvantage of using Metaverse technology in HR management was determined to be the risk of disconnecting people from reality, with one in every two participants choosing this. Half of the participants see Metaverse technology's reduction of human interaction and data privacy and security concerns as a disadvantage. The rate of those who say that it may cause people to have psychological problems is approximately 40\%.
In conclusion, this study has provided valuable insights into the perceptions of participants regarding the integration of emerging technologies, such as artificial intelligence, virtual reality, augmented reality, and Metaverse, into human resources management processes. The findings suggest that there is a significant interest among participants in leveraging these technologies to enhance various aspects of HR management, including recruitment, training, and performance evaluation.
While there is a general optimism about the benefits of these technologies, such as increased efficiency and improved candidate evaluation, participants also acknowledge potential challenges and drawbacks, such as ethical concerns, data security risks, and the risk of disconnecting from reality.
Overall, this study highlights the importance of carefully considering both the opportunities and challenges associated with the adoption of emerging technologies in HR management. Future research and implementation efforts should focus on addressing these concerns while maximizing the potential benefits to create more effective and sustainable HR practices in the digital age.

\section*{Acknowledgment}
The study is carried out under the decision of the Manisa Celal Bayar University Science and Engineering Sciences Scientific Research and Publication Ethics Board, designated as number 1, made at Meeting Number 2023/04 dated 15.09.2023. We thank Manisa Celal Bayar University for their support.

\section*{Authors' Contribution}
Ö.A. participated in writing the article, as well as in the creation and implementation of the survey, and the processes of data acquisition and analysis. E.K. contributed to the writing of the article, as well as to the survey processes and evaluations. N.G.N. contributed to the evaluation of the survey results in writing the article. All authors read and approved the final version of the manuscript.

\section*{Conflict of Interest}
The author(s) declare that they have no conflicting interests.

\section*{Funding}
This research did not receive any outside funding or support. The authors report no involvement in the research by the sponsor that could have influenced the outcome of this work.

\section*{Data Availability}
The data used in the article can be used for scientific purposes and shared with anyone interested in compliance with ethical principles. To share the data, it is necessary to contact the corresponding author of the article.

\end{document}